 \documentclass[pmlr,twocolumn]{jmlr} % W&CP article

\usepackage{booktabs}
\usepackage[load-configurations=version-1]{siunitx} % newer version

\usepackage{listings}
\NewDocumentCommand{\codeword}{v}{%
\texttt{\textcolor{brown}{#1}}%
}

\theorembodyfont{\upshape}
\theoremheaderfont{\scshape}
\theorempostheader{:}
\theoremsep{\newline}

\jmlrvolume{ML4H Extended Abstract Arxiv Index}
\jmlryear{2020}
\jmlrsubmitted{2020}
\jmlrpublished{}
\jmlrworkshop{Machine Learning for Health (ML4H) 2020}

\title[Expression of Distinct Immune Pathways for SLE Patients]{Stratification of Systemic Lupus Erythematosus Patients Using Gene Expression Data to Reveal Expression of Distinct Immune Pathways}

\author{%
\Name{Aditi Deokar} \Email{adeokar@bu.edu}\\
\addr Boston University Academy, Boston, MA, USA
}

\begin{document}

\maketitle

\begin{abstract}
Systemic lupus erythematosus (SLE) is the tenth leading cause of death in females 15-24 years old in the US. The diversity of symptoms and immune pathways expressed in SLE patients causes difficulties in treating SLE as well as in new clinical trials. This study used unsupervised learning on gene expression data from adult SLE patients to separate patients into clusters. The dimensionality of the gene expression data was reduced by three separate methods (PCA, UMAP, and a simple linear autoencoder) and the results from each of these methods were used to separate patients into six clusters with k-means clustering. 

The clusters revealed three separate immune pathways in the SLE patients that caused SLE. These pathways were: (1) high interferon levels, (2) high autoantibody levels, and (3) dysregulation of the mitochondrial apoptosis pathway. The first two pathways have been extensively studied in SLE. However, mitochondrial apoptosis has not been investigated before to the best of our knowledge as a standalone cause of SLE, independent of autoantibody production, indicating that mitochondrial proteins could lead to a new set of therapeutic targets for SLE in future research.
\end{abstract}

\section{Introduction}
\label{sec:intro}

Systemic lupus erythematosus (SLE) is the tenth most common cause of death among females 15-24 years old in the US \citep{Yen2018}. SLE is one of many autoimmune diseases, which are diseases in which a patient’s immune system mistakes parts of their own body as foreign, attacking their healthy organs and tissue \citep{LupusFoundationofAmerica2020}. 

SLE can be driven by defects in the innate immune system and/or the adaptive immune system. SLE patients are often characterized by high levels of interferon-1, which causes inflammation in the innate immune system in response to viruses. In SLE, high interferon levels can be caused by a variety of factors, such as neutrophil extracellular traps \citep{Bengtsson2017}. Most SLE patients also have high levels of autoantibodies, which are antibodies directed against self cells and are created by mature B cells (plasma cells) \citep{Dema2016}. Autoantibodies cause a much more targeted response than the innate immune system, but SLE patients can have a wide range of autoantibodies - one study found over 180 autoantibodies expressed in SLE patients \citep{Yaniv2015}. Some patients with lupus do not even have autoantibodies, and many of the autoantibodies in SLE are also found in other rheumatic diseases \citep{Egner2000}.

The heterogeneity of lupus symptoms and immune pathways affected makes it difficult to treat, because different drugs work well on different patients. \citet{Merrill2017} found that certain standard drugs (anti-rheumatic drugs and immunosuppressants) affect immune pathways differently in interferon-low and interferon-high patients. While there is still debate on whether SLE is one disease or many \citep{Agmon-Levin2012}, it is clear that subdividing SLE patients into categories will help treat patients. 

Previous studies have tackled this problem by dividing patients based on antibody levels \citep{Artim-Esen2014}, gene expression \citep{Toro-Dominguez2018}, and immune molecule levels \citep{Hamilton2018}. However, none of these studies have reached a consensus on the best subdivision of SLE. \citet{Guthridge2020} used all three of these factors to divide SLE patients into seven clusters with unsupervised machine learning. However, in practice, gathering these different types of patient data to categorize a patient into an SLE subdivision is infeasible. 

In this study, we use unsupervised machine learning to categorize SLE patients using only gene expression data, which is more accessible than all three types of data combined. This would help determine if gene expression data alone reveals similar patterns in immune pathway expression as does its combination with antibody levels and immune molecule levels.

\section{Data and Methods}
\label{sec:meth}

We use a gene expression dataset available on GEO (accession number GSE138458) containing data collected by ~\citet{Guthridge2020}. The data includes 336 samples in total, with 24 control patients and 198 SLE patients. 108 of the SLE patients have two or more samples taken. The data analysis methods are shown in \figureref{fig:flowchart} in \ref{flowchart}. Data pre-normalized by ~\citet{Guthridge2020} is used, employing bgAdjust background correction, vst variance stabilizing transformation, and rank invariant normalization, and outlier removal (1 control and 5 SLE).

Given the high dimensionality nature of gene expression data, dimension reduction techniques are required. Prior work has used unsupervised learning following dimension reduction of gene expression data for other diseases such as cancer \citep{Shi2010}. The dimensionality of our 47,323 gene data is reduced using three separate methods: \emph{Principal Component Analysis} (PCA), \emph{Uniform Manifold Approximation and Projection} (UMAP), and a simple \emph{autoencoder} (AE), which are intended to minimize the effects of random variation on the unsupervised clustering model in different ways.

With both PCA and UMAP, 200 reduced features are selected. In PCA, these explain 96.29\% of the variance in the original 47,323 genes. UMAP is a nonlinear model (unlike PCA) similar to t-SNE, used for visualization as well as nonlinear dimension reduction \citep{McInnes2018}. The autoencoder aims to reduce the loss of information between the original inputs (genes) and the decoded output of the same dimension. AEs with both linear and sigmoid activation functions are validated and the linear AE is found to perform much better after 100 epochs (validation loss 0.068) than the sigmoid AE (validation loss 48.28). So, we select the 1000 encoded components from the linear AE for subsequent clustering.

The three datasets with reduced features are then used for k-means clustering. To determine the best number of clusters, we use the YellowBrick Python package, which creates visualizations quantifying the ``elbow method'' used on the metrics distortion score, silhouette score, and Calinski-Harabasz score. All these metrics are found to converge on 6 clusters for each dataset. k-means clustering is then used to derive 6 clusters from each dataset.

For visualization and interpretation of the clusters, we use 27 pre-existing modules created by \citet{Chaussabel2008}. Each module represents a group of genes with a common function. These are used to calculate module scores for the three datasets. Module scores for each cluster represent the percentage of genes in each module that were significantly upregulated (i.e., overexpressed) or downregulated (i.e., underexpressed) in that cluster as compared to the controls, based on a two-tailed t-test ($p < 0.05$).

\section{Results and Discussion}
\label{sec:res}

\begin{figure}[h]
\floatconts
    {fig:heatmaps}
    {\caption{Comparative gene underexpression or overexpression for gene expression modules across clusters.}}
    {%
     \subfigure[PCA heatmap]{\label{fig:PCA heatmap}%
         \includegraphics[width=1.0\linewidth]{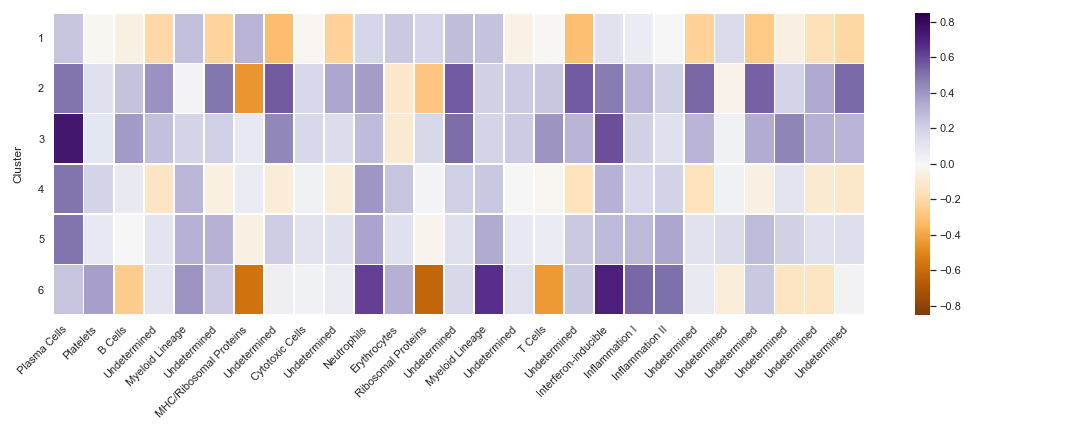}}%
         \qquad
     \subfigure[UMAP heatmap]{\label{fig:UMAP heatmap}%
         \includegraphics[width=1.0\linewidth]{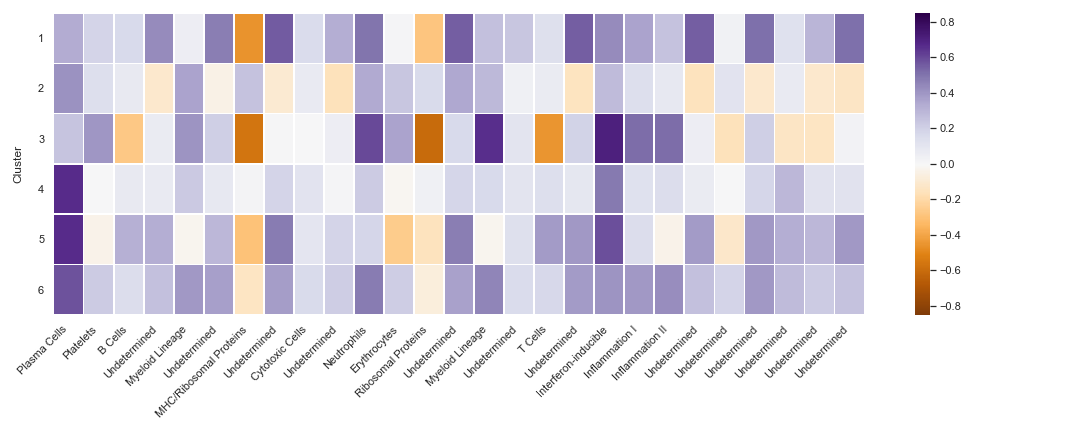}}%
         \qquad
     \subfigure[Autoencoder heatmap]{\label{fig:Autoencoder heatmap}%
         \includegraphics[width=1.0\linewidth]{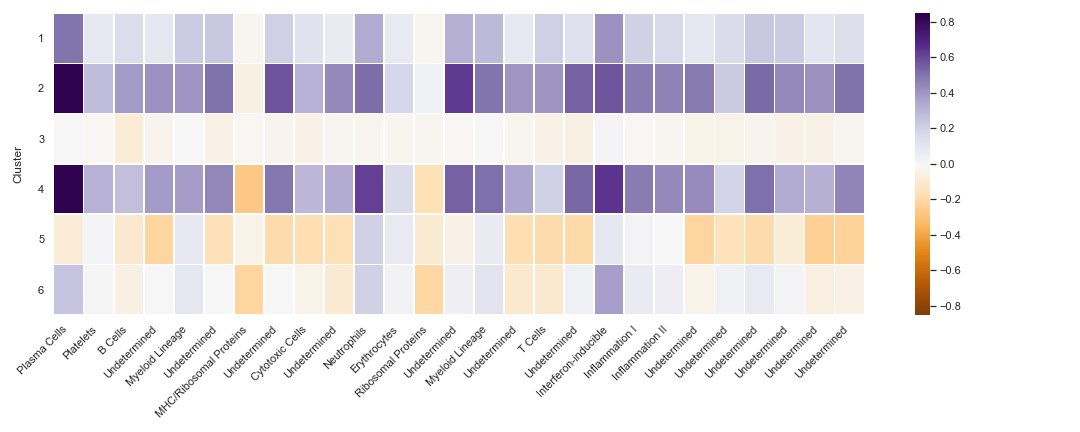}}
    }
\end{figure}

\figureref{fig:heatmaps} shows heatmaps generated from the module scores for the 3 feature-reduced datasets. These heatmaps show the percentage of underexpressed (brown) or overexpressed (purple) genes for SLE patients as compared to the controls. 

The clusters originating from the PCA and UMAP dimensionality reductions show very similar patterns in the upregulated and downregulated modules, while the clusters originating from the AE mostly show a consistent level of increased or decreased gene expression across all modules, excepting cluster 5. 

The patients in the clusters created from the PCA and UMAP dimensionality reduction techniques and AE cluster 5 can be designated as belonging to one of three groups: (a) interferon-driven SLE, (b) autoantibody-driven SLE, and (c) SLE caused by mitochondrial apoptosis. The first two groups substantiate results from prior literature, while the third group presents a pathway that suggests a novel cause of SLE.

\subsection{Interferon-driven SLE}
\label{subsec:inter}

In lupus, type 1 interferon levels are often elevated, which can lead to inflammation and tissue damage caused by the innate immune system \citep{Crow2014}. PCA cluster 6 and UMAP cluster 3 in \figureref{fig:heatmaps}
both display substantial upregulation of genes related to interferons and inflammation. These two clusters validate the patterns also observed in \citet{Guthridge2020}'s clusters 1, 4 and 6. All of these upregulated genes are related to the innate immune response. Those patients also have underexpressed B and T cells and normal expression of plasma cells, which would all be overexpressed if production of autoantibodies by plasma cells was the main reason for autoimmunity, rather than interferon levels.

\subsection{Autoantibody-driven SLE}
\label{subsec:anti}

Many of the other PCA and UMAP clusters displayed upregulation of antibody-producing plasma cells; particularly PCA clusters 2, 3, 4 and 5 and UMAP clusters 4, 5, and 6. \citet{Guthridge2020} observed a similar trend, where their clusters 2, 3, and 5 had higher T cell, B cell, and plasma cell related expression. While autoantibodies are known to be common in SLE, the diversity of autoantibodies (as discussed in \citet{Yaniv2015}) means that there is still work to be done understanding what is different among the antibodies produced in these four PCA clusters and three UMAP clusters. Some of these differences might come from genes used to create the clusters that were not included in the modules used for the heatmap visualization.

\citet{Brant2020}, who grouped lupus patients based on their correlation between gene expression and disease activity, found one cluster where neutrophil levels correlated to disease activity and one where lymphocyte levels correlated to disease activity. Since neutrophil extracellular traps are one way that interferon levels become elevated, their neutrophil-correlated group might correspond to our high-interferon group, and their lymphocyte-correlated group might correspond to our antibody-driven group. More analysis should be done on disease activity correlation in our data to confirm this.

\subsection{SLE caused by mitochondrial apoptosis}
\label{subsec:apop}

PCA clusters 1 and 4, UMAP cluster 2, and Autoencoder cluster 5 display a different pattern from many of the other clusters. In these clusters, many of the modules labeled as \emph{Undetermined} by \citet{Chaussabel2008} were underexpressed. A closer look at the genes in these \emph{Undetermined} modules reveals that they include mitochondrial ribosomal proteins, mitochondrial elongation factors, and proteins in the cAMP-signaling pathway. Mitochondrial ribosomal proteins, in addition to their ribosomal functions, are involved in apoptotic (programmed cell death) pathways \citep{Kim2017}, and cAMP signaling regulates mitochondrial apoptosis \citep{Valsecchi2013}. Apoptosis is known to be a factor in SLE, but mainly because ineffective clearance of apoptotic cells can expose B and T cells to intracellular material, leading to the creation of autoantibodies against this intracellular material \citep{Mevorach2003}.

We suggest that for the patients in these clusters, dysregulation of mitochondrial pathways or signaling from outside molecules, possibly lymphocytes, could cause mitochondrial apoptotic pathways to become activated in healthy cells, destroying healthy cells as is characteristic of SLE. These healthy cells would have a range of gene expression of mitochondrial proteins, but the cells with higher expression of the proteins would activate the apoptotic pathway and die. Only cells with lower expression levels would survive, so lower expression levels were found in our study. These lower expression levels would also impair mitochondrial functions, which has been observed to be true in SLE patients \citep{Leishangthem2016}.

Cluster 7 from the \citet{Guthridge2020} study also had low expression of mitochondrial respiration and mitochondrial stress genes (not discussed in their study). The discovery of this cluster of patients using two completely different machine learning approaches corroborates the idea that the mitochondrial apoptotic pathway is a novel cause for SLE. Future studies should investigate to a further extent the mitochondrial apoptotic pathway in SLE patients as a reason for destruction of self cells in addition to a way that autoantibodies are produced.

\bibliography{ms}

\begin{thebibliography}{23}
\providecommand{\natexlab}[1]{#1}
\providecommand{\url}[1]{\texttt{#1}}
\expandafter\ifx\csname urlstyle\endcsname\relax
  \providecommand{\doi}[1]{doi: #1}\else
  \providecommand{\doi}{doi: \begingroup \urlstyle{rm}\Url}\fi

\bibitem[Agmon-Levin et~al.(2012)Agmon-Levin, Mosca, Petri, and
  Shoenfeld]{Agmon-Levin2012}
N.~Agmon-Levin, M.~Mosca, M.~Petri, and Y.~Shoenfeld.
\newblock {Systemic lupus erythematosus one disease or many?}
\newblock \emph{Autoimmunity Reviews}, 11\penalty0 (8):\penalty0 593--595,
  2012.
\newblock ISSN 15689972.
\newblock \doi{10.1016/j.autrev.2011.10.020}.
\newblock URL \url{http://dx.doi.org/10.1016/j.autrev.2011.10.020}.

\bibitem[Artim-Esen et~al.(2014)Artim-Esen, {\c{C}}ene, Şahinkaya, Ertan,
  Pehlivan, Kamali, G{\"{u}}l, {\"{O}}cal, Aral, and
  Inan{\c{c}}]{Artim-Esen2014}
Bahar Artim-Esen, Erhan {\c{C}}ene, Yasemin Şahinkaya, Semra Ertan,
  {\"{O}}zlem Pehlivan, Sevil Kamali, Ahmet G{\"{u}}l, Lale {\"{O}}cal, Orhan
  Aral, and Murat Inan{\c{c}}.
\newblock {Cluster analysis of autoantibodies in 852 patients with systemic
  lupus erythematosus from a single center}.
\newblock \emph{Journal of Rheumatology}, 41\penalty0 (7):\penalty0 1304--1310,
  2014.
\newblock ISSN 14992752.
\newblock \doi{10.3899/jrheum.130984}.

\bibitem[Bengtsson and R{\"{o}}nnblom(2017)]{Bengtsson2017}
Anders~A. Bengtsson and Lars R{\"{o}}nnblom.
\newblock {Role of interferons in SLE}.
\newblock \emph{Best Practice and Research: Clinical Rheumatology}, 31\penalty0
  (3):\penalty0 415--428, 2017.
\newblock ISSN 15321770.
\newblock \doi{10.1016/j.berh.2017.10.003}.

\bibitem[Brant et~al.(2020)Brant, Rietman, Klement, Cavaglia, and
  Tuszynski]{Brant2020}
Elizabeth~J. Brant, Edward~A. Rietman, Giannoula~Lakka Klement, Marco Cavaglia,
  and Jack~A. Tuszynski.
\newblock {Personalized therapy design for systemic lupus erythematosus based
  on the analysis of protein-protein interaction networks}.
\newblock \emph{PLoS ONE}, 15\penalty0 (3):\penalty0 1--16, 2020.
\newblock ISSN 19326203.
\newblock \doi{10.1371/journal.pone.0226883}.
\newblock URL \url{http://dx.doi.org/10.1371/journal.pone.0226883}.

\bibitem[Chaussabel et~al.(2008)Chaussabel, Quinn, Shen, Patel, Glaser,
  Baldwin, Stichweh, Blankenship, Li, Munagala, Bennett, Allantaz, Mejias,
  Ardura, Kaizer, Monnet, Allman, Randall, Johnson, Lanier, Punaro, Wittkowski,
  White, Fay, Klintmalm, Ramilo, Palucka, Banchereau, and
  Pascual]{Chaussabel2008}
Damien Chaussabel, Charles Quinn, Jing Shen, Pinakeen Patel, Casey Glaser,
  Nicole Baldwin, Dorothee Stichweh, Derek Blankenship, Lei Li, Indira
  Munagala, Lynda Bennett, Florence Allantaz, Asuncion Mejias, Monica Ardura,
  Ellen Kaizer, Laurence Monnet, Windy Allman, Henry Randall, Diane Johnson,
  Aimee Lanier, Marilynn Punaro, Knut~M. Wittkowski, Perrin White, Joseph Fay,
  Goran Klintmalm, Octavio Ramilo, A.~Karolina Palucka, Jacques Banchereau, and
  Virginia Pascual.
\newblock {A Modular Analysis Framework for Blood Genomics Studies: Application
  to Systemic Lupus Erythematosus}.
\newblock \emph{Immunity}, 29\penalty0 (1):\penalty0 150--164, 2008.
\newblock ISSN 10747613.
\newblock \doi{10.1016/j.immuni.2008.05.012}.

\bibitem[Crow(2014)]{Crow2014}
Mary~K. Crow.
\newblock {Type I Interferon in the Pathogenesis of Lupus}.
\newblock \emph{The Journal of Immunology}, 192\penalty0 (12):\penalty0
  5459--5468, 2014.
\newblock ISSN 0022-1767.
\newblock \doi{10.4049/jimmunol.1002795}.

\bibitem[Dema and Charles(2016)]{Dema2016}
Barbara Dema and Nicolas Charles.
\newblock {Autoantibodies in SLE: Specificities, Isotypes and Receptors}.
\newblock \emph{Antibodies}, 5\penalty0 (1):\penalty0 2, 2016.
\newblock ISSN 2073-4468.
\newblock \doi{10.3390/antib5010002}.

\bibitem[Egner(2000)]{Egner2000}
William Egner.
\newblock {The use of laboratory tests in the diagnosis of SLE}.
\newblock \emph{Journal of Clinical Pathology}, 53\penalty0 (6):\penalty0
  424--432, 2000.
\newblock ISSN 00219746.
\newblock \doi{10.1136/jcp.53.6.424}.

\bibitem[Guthridge et~al.(2020)Guthridge, Lu, Tran, Arriens, Aberle, Kamp,
  Munroe, Dominguez, Gross, DeJager, Macwana, Bourn, Apel, Thanou, Chen,
  Chakravarty, Merrill, and James]{Guthridge2020}
Joel~M. Guthridge, Rufei Lu, Ly~Thi~Hai Tran, Cristina Arriens, Teresa Aberle,
  Stan Kamp, Melissa~E. Munroe, Nicolas Dominguez, Timothy Gross, Wade DeJager,
  Susan~R. Macwana, Rebecka~L. Bourn, Stephen Apel, Aikaterini Thanou, Hua
  Chen, Eliza~F. Chakravarty, Joan~T. Merrill, and Judith~A. James.
\newblock {Adults with systemic lupus exhibit distinct molecular phenotypes in
  a cross-sectional study}.
\newblock \emph{EClinicalMedicine}, 20:\penalty0 100291, 2020.
\newblock ISSN 25895370.
\newblock \doi{10.1016/j.eclinm.2020.100291}.
\newblock URL \url{https://doi.org/10.1016/j.eclinm.2020.100291}.

\bibitem[Hamilton et~al.(2018)Hamilton, Wu, Yang, Luo, Liu, Li, {L.
  Mattheyses}, Sanz, Chatham, Hsu, and Mountz]{Hamilton2018}
Jennie~A. Hamilton, Qi~Wu, PingAr Yang, Bao Luo, Shanrun Liu, Jun Li, Alexa {L.
  Mattheyses}, Ignacio Sanz, W.~Winn Chatham, Hui-Chen Hsu, and John~D. Mountz.
\newblock {Cutting Edge: Intracellular IFN-$\beta$ and Distinct Type I IFN
  Expression Patterns in Circulating Systemic Lupus Erythematosus B Cells}.
\newblock \emph{The Journal of Immunology}, 201\penalty0 (8):\penalty0
  2203--2208, 2018.
\newblock ISSN 0022-1767.
\newblock \doi{10.4049/jimmunol.1800791}.

\bibitem[Kegerreis et~al.(2019)Kegerreis, Catalina, Bachali, Geraci, Labonte,
  Zeng, Stearrett, Crandall, Lipsky, and Grammer]{Kegerreis2019}
Brian Kegerreis, Michelle~D. Catalina, Prathyusha Bachali, Nicholas~S. Geraci,
  Adam~C. Labonte, Chen Zeng, Nathaniel Stearrett, Keith~A. Crandall, Peter~E.
  Lipsky, and Amrie~C. Grammer.
\newblock {Machine learning approaches to predict lupus disease activity from
  gene expression data}.
\newblock \emph{Scientific Reports}, 9\penalty0 (1):\penalty0 1--12, 2019.
\newblock ISSN 20452322.
\newblock \doi{10.1038/s41598-019-45989-0}.
\newblock URL \url{http://dx.doi.org/10.1038/s41598-019-45989-0}.

\bibitem[Kim et~al.(2017)Kim, Maiti, and Barrientos]{Kim2017}
Hyun-Jung Kim, Priyanka Maiti, and Antoni Barrientos.
\newblock {Mitochondrial ribosomes in cancer}.
\newblock \emph{Seminars in Cancer Biology}, 47\penalty0 (3):\penalty0 67--81,
  dec 2017.
\newblock ISSN 1044579X.
\newblock \doi{10.1016/j.semcancer.2017.04.004}.
\newblock URL
  \url{https://linkinghub.elsevier.com/retrieve/pii/S1044579X17300962}.

\bibitem[Leishangthem et~al.(2016)Leishangthem, Sharma, and
  Bhatnagar]{Leishangthem2016}
B.~D. Leishangthem, A.~Sharma, and Archana Bhatnagar.
\newblock {Role of altered mitochondria functions in the pathogenesis of
  systemic lupus erythematosus}.
\newblock \emph{Lupus}, 25\penalty0 (3):\penalty0 272--281, 2016.
\newblock ISSN 14770962.
\newblock \doi{10.1177/0961203315605370}.

\bibitem[{Lupus Foundation of America}(2020)]{LupusFoundationofAmerica2020}
{Lupus Foundation of America}.
\newblock {What is lupus?}, 2020.
\newblock URL \url{https://www.lupus.org/resources/what-is-lupus}.

\bibitem[McInnes et~al.(2020)McInnes, Healy, and Melville]{McInnes2018}
Leland McInnes, John Healy, and James Melville.
\newblock {UMAP: Uniform Manifold Approximation and Projection for Dimension
  Reduction}, 2020.
\newblock URL \url{http://arxiv.org/abs/1802.03426}.

\bibitem[Merrill et~al.(2017)Merrill, Immermann, Whitley, Zhou, Hill, O'Toole,
  Reddy, Honczarenko, Thanou, Rawdon, Guthridge, James, and
  Sridharan]{Merrill2017}
Joan~T. Merrill, Fred Immermann, Maryann Whitley, Tianhui Zhou, Andrew Hill,
  Margot O'Toole, Padmalatha Reddy, Marek Honczarenko, Aikaterini Thanou, Joe
  Rawdon, Joel~M. Guthridge, Judith~A. James, and Sudhakar Sridharan.
\newblock {The Biomarkers of Lupus Disease Study: A Bold Approach May Mitigate
  Interference of Background Immunosuppressants in Clinical Trials}.
\newblock \emph{Arthritis and Rheumatology}, 69\penalty0 (6):\penalty0
  1257--1266, 2017.
\newblock ISSN 23265205.
\newblock \doi{10.1002/art.40086}.

\bibitem[Mevorach(2003)]{Mevorach2003}
Dror Mevorach.
\newblock {Systemic Lupus Erythematosus and Apoptosis}.
\newblock \emph{Clinical reviews in allergy {\&} immunology}, 25:\penalty0
  49--59, 2003.

\bibitem[Petri et~al.(2019)Petri, Fu, Ranger, Allaire, Cullen, Magder, and
  Zhang]{Petri2019}
Michelle Petri, Wei Fu, Ann Ranger, Norm Allaire, Patrick Cullen, Laurence~S.
  Magder, and Yuji Zhang.
\newblock {Association between changes in gene signatures expression and
  disease activity among patients with systemic lupus erythematosus}.
\newblock \emph{BMC Medical Genomics}, 12\penalty0 (1):\penalty0 1--9, 2019.
\newblock ISSN 17558794.
\newblock \doi{10.1186/s12920-018-0468-1}.

\bibitem[Shi and Luo(2010)]{Shi2010}
Jinlong Shi and Zhigang Luo.
\newblock Nonlinear dimensionality reduction of gene expression data for
  visualization and clustering analysis of cancer tissue samples.
\newblock \emph{Computers in Biology and Medicine}, 40\penalty0 (8):\penalty0
  723 -- 732, 2010.
\newblock ISSN 0010-4825.
\newblock \doi{https://doi.org/10.1016/j.compbiomed.2010.06.007}.
\newblock URL
  \url{http://www.sciencedirect.com/science/article/pii/S0010482510000958}.

\bibitem[Toro-Dom{\'{i}}nguez et~al.(2018)Toro-Dom{\'{i}}nguez,
  Martorell-Marug{\'{a}}n, Goldman, Petri, Carmona-S{\'{a}}ez, and
  Alarc{\'{o}}n-Riquelme]{Toro-Dominguez2018}
Daniel Toro-Dom{\'{i}}nguez, Jordi Martorell-Marug{\'{a}}n, Daniel Goldman,
  Michelle Petri, Pedro Carmona-S{\'{a}}ez, and Marta~E.
  Alarc{\'{o}}n-Riquelme.
\newblock {Stratification of Systemic Lupus Erythematosus Patients Into Three
  Groups of Disease Activity Progression According to Longitudinal Gene
  Expression}.
\newblock \emph{Arthritis and Rheumatology}, 70\penalty0 (12):\penalty0
  2025--2035, 2018.
\newblock ISSN 23265205.
\newblock \doi{10.1002/art.40653}.

\bibitem[Valsecchi et~al.(2013)Valsecchi, Ramos-Espiritu, Buck, Levin, and
  Manfredi]{Valsecchi2013}
Federica Valsecchi, Lavoisier~S. Ramos-Espiritu, Jochen Buck, Lonny~R. Levin,
  and Giovanni Manfredi.
\newblock {cAMP and mitochondria}.
\newblock \emph{Physiology}, 28\penalty0 (3):\penalty0 199--209, 2013.
\newblock ISSN 15489213.
\newblock \doi{10.1152/physiol.00004.2013}.

\bibitem[Yaniv et~al.(2015)Yaniv, Twig, Shor, Furer, Sherer, Mozes, Komisar,
  Slonimsky, Klang, Lotan, Welt, Marai, Shina, Amital, and
  Shoenfeld]{Yaniv2015}
Gal Yaniv, Gilad Twig, Dana Ben~Ami Shor, Ariel Furer, Yaniv Sherer, Oshry
  Mozes, Orna Komisar, Einat Slonimsky, Eyal Klang, Eyal Lotan, Mike Welt,
  Ibrahim Marai, Avi Shina, Howard Amital, and Yehuda Shoenfeld.
\newblock {A volcanic explosion of autoantibodies in systemic lupus
  erythematosus: A diversity of 180 different antibodies found in SLE
  patients}.
\newblock \emph{Autoimmunity Reviews}, 14\penalty0 (1):\penalty0 75--79, 2015.
\newblock ISSN 18730183.
\newblock \doi{10.1016/j.autrev.2014.10.003}.
\newblock URL \url{http://dx.doi.org/10.1016/j.autrev.2014.10.003}.

\bibitem[Yen and Singh(2018)]{Yen2018}
Eric~Y Yen and Ram~R Singh.
\newblock {Lupus – An Unrecognized Leading Cause of Death in Young Women:
  Population-based Study Using Nationwide Death Certificates, 2000–2015}.
\newblock \emph{Arthritis and Rheumatology}, 70\penalty0 (8):\penalty0
  1251--1255, 2018.
\newblock \doi{10.1002/art.40512}.

\end{thebibliography}

\appendix

\section{Supplement}

\subsection{Methods Flowchart}
\label{flowchart}

\begin{figure}[htbp]
    \includegraphics[width=1.0\linewidth]{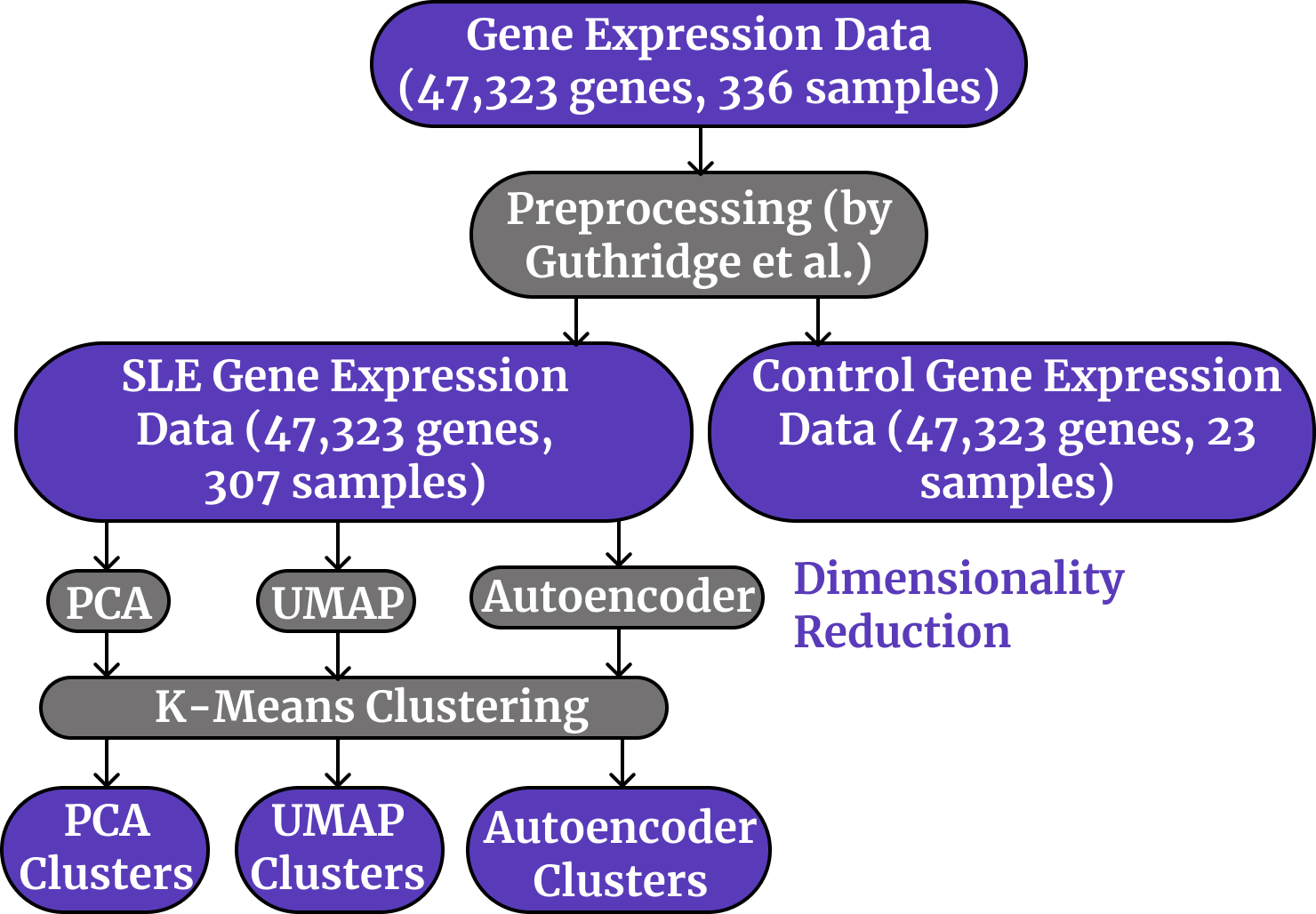}
    \caption{Overview of data analysis}
    \label{fig:flowchart}
\end{figure}

 \figureref{fig:flowchart} illustrates the data analysis steps undertaken in this study. A detailed description is provided in Section \ref{sec:meth}.

\subsection{Dimensionality Reduction Models' Specifications}
\label{sec:note-3}
The following specifications were used in the three dimensionality reduction techniques.
\begin{itemize}
    \item \emph{Autoencoder}: \codeword{keras} API was used for the simple autoencoder with: Input dimension: \codeword{(47323,)}, Output dimension: \codeword{(1000,)}, \codeword{activation = `linear'} in encoded and decoded layers, \codeword{epochs = 50}, \codeword{optimizer = `adam'}, \codeword{loss = `mse'}, \codeword{batch_size = 64}, \codeword{shuffle = True}, \codeword{validation_split = 0.2}.
    \item \emph{UMAP}: \codeword{UMAP} parameters used: \codeword{n_components = 200}, \codeword{n_neighbors = 15}, \codeword{min_dist = 0.1}, \codeword{metric = `euclidean'}.
    \item \emph{PCA}: Linear dimensionality reduction using Singular Value Decomposition (SVD) was used with the PCA class in \codeword{sklearn} API with the following parameters:  \codeword{n_components = 200}, \codeword{svd_solver = `randomized'}.
\end{itemize}

\subsection{Analyzing Patients with Multiple Samples}
\label{sec:note-1}

\begin{figure}[h]
\floatconts
    {fig:same/diff}
    {\caption{Patients in each cluster, of those who had multiple samples taken, who were put in the same cluster or different cluster for (a) PCA clusters, (b) UMAP clusters, (c) Autoencoder clusters.}}
    {%
     \subfigure[PCA clusters]{\label{fig:PCA same/diff}%
         \includegraphics[width=0.65\linewidth]{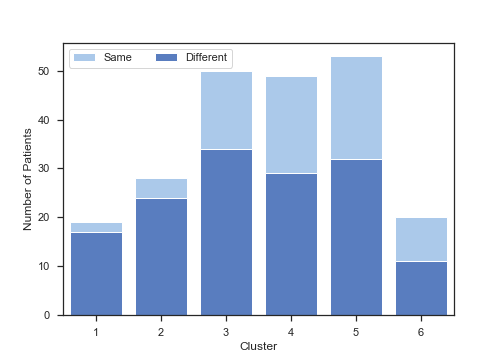}}%
         \qquad
     \subfigure[UMAP clusters]{\label{fig:UMAP same/diff}%
         \includegraphics[width=0.65\linewidth]{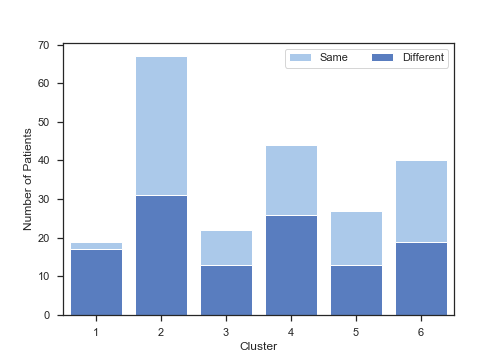}}%
         \qquad
     \subfigure[Autoencoder clusters]{\label{fig:Autoencoder same/diff}%
         \includegraphics[width=0.65\linewidth]{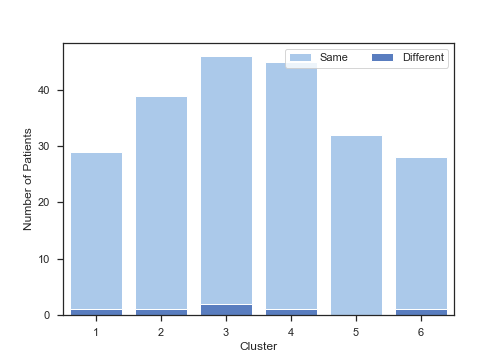}}
    }
\end{figure}
For patients who had multiple samples taken, k-means following the autoencoder classified them into the same cluster 97.3\% of the time, while k-means following PCA and UMAP classified them into the same cluster 32.9\% and 45.7\% of the time respectively (\figureref{fig:same/diff}). While gene expression data is correlated with SLE disease activity \citep{Kegerreis2019, Toro-Dominguez2018}, \citet{Petri2019} found that the majority of gene expression signatures were stable in patients over time. This suggests that the autoencoder's dimensionality reduction may have emphasized the stable gene expression signatures, causing them to be a major factor in the clustering, but that PCA and UMAP, which aimed to preserve more of the variance in the data, did not maintain the data from genes whose expression was stable over time. Many of these more stable genes might not have been related to the immune system, so they were not included in ~\citet{Chaussabel2008}'s coexpression modules. Thus, the more variable modules that were in the heatmap would have shown a lot of variation between the patients in each cluster, causing the clusters in the autoencoder heatmap to show a more consistent level of expression across all genes in the modules. Further analysis should be done to determine the level of variation in gene expression in the modules for the autoencoder clusters in comparison to the PCA and UMAP clusters, and to determine whether the PCA and UMAP clusters correlated to disease activity more than the autoencoder clusters, which this idea would imply.

\end{document}